# Primary breakup of liquid jet– Effect of jet velocity profile

Balaji Srinivasan and Anubhav Sinha*

Department of Mechanical Engineering

Indian Institute of Technology (Banaras Hindu University)

Varanasi, India – 221005

* Corresponding author email: er.anubhav@gmail.com

## Abstract

*The present work examines the effect of the velocity profile on primary breakup of liquid jets emanating from fuel injectors. Direct Numerical Simulation (DNS) is used to simulate liquid jet breakup. Different velocity profiles are imposed on the liquid and their effect on breakup is examined. It is a common practice in the literature to use flat or uniform velocity profiles in such studies. The validity of this assumption is assessed and its implications are highlighted. Droplet sizes and degree of atomization are compared for all the cases. Further, a detailed comparison of jet breakup structure is made for two cases – parabolic and power law velocity profiles. The liquid surface is observed to show two-dimensional waves initially, which subsequently transform into three-dimensional waves and give rise to ligament formation and surface breakup. Tip vortex rollup and its role in jet breakup is discussed. The distinction between different velocity profiles is examined in detail in terms of surface waves, degree of atomization, and jet structure.*

## Introduction

Understanding primary atomization is crucial for injector design. The primary atomization of a liquid jet can be investigated using experimental, numerical as well as analytical methods. However, non-intrusive experiments are challenging to conduct. Optical methods (Linne, 2013) can be employed, but breakup commences very close to the nozzle producing a large number of small droplets that obstruct imaging. Further, the aerodynamic interaction between the liquid jet and surrounding gas plays a critical role in governing breakup (Sinha, 2019), which is challenging to capture accurately in numerical simulations (Gorokhovski and Hermann, 2008). Historically, computational studies have approximated jet breakup using empirical relations and simplistic models (Reitz, 2004; Wu et al., 2021). Recent improvement in computational power has opened new opportunities to probe this configuration in detail. However, LES and RANS-based approaches cannot provide a true representation of irregularities on the jet surface. Hence, DNS appears to be the most suitable option to aid in understanding the primary breakup of liquid jets. To capture the true physics using DNS, boundary conditions need to be correctly defined. As explained in the following paragraphs, defining jet inlet conditions, especially velocity profile is critical to model breakup but has often been overlooked. Physically, the liquid inside the nozzle maintains no-slip condition at the wall. As soon as liquid emerges out of the injector tube, the no-slip boundary is suddenly removed. This causes a redistribution in velocity which is believed to significantly contribute to surface instability (Park and Heister, 2006; Sinha, 2019).

Jarrahbashi et al. (2014, 2016) carried out DNS simulations of liquid jet breakup using the level-set method. Their focus is on early breakup structures like holes, ligaments, and bridge formations. They have used an exponential velocity profile as a boundary condition for the liquid jet. The results reveal rich structural aspects of round jet breakup. Another study was carried out for planar sheets (Zandian et al., 2017) where similar structures were observed. However, the gas-to-liquid density ratio is much higher than usual in engineering applications. Lebas et al. (2009) investigated the primary breakup of diesel jets in evaporating conditions and reported a reasonable match with their model and DNS study. Ling et al. (2017) have used a VOF solver to compare diesel and biodiesel atomization behavior. They have reported diesel jets to produce more droplets than biodiesel jets. They have attributed the higher viscosity of biodiesel for the difference in atomization behavior. Shinjo and Umemura (2011a, b) investigated the primary breakup of a diesel jet using DNS results. They have employed a level-set method for interface tracking. They ignore the effect of injector geometry and velocity profile and have used a uniform velocity profile as the jet inlet boundary condition. Their results show the dominant role of the tip vortex in causing instability in the jet surface and subsequent breakup. They have also probed the wave growth, ligament, and droplet formation in detail (Shinjo and Umemura, 2010) and have identified surface features resulting in ligament formation and rupture.

Popinet (2003) has developed a VOF solver Gerris, which utilizes adaptive mesh refinement. The solver shows second-order convergence in spatial and temporal domains. Fuster et al. (2009) have demonstrated the applicability of Gerris in the primary atomization of a liquid jet. Basilisk is a successor of Gerris (Popinet, 2015). Pairetti et al. (2020) have explored the effect of mesh resolution in liquid jet breakup using Basilisk. They have observed mesh resolution to impact breakup and have provided useful guidelines to account for this factor while selecting mesh size. Gerris is also used to investigate the primary breakup of gasoline jets for direct injection under non-evaporating conditions (Zhang et al., 2020). The paper explores a non-symmetric injector and its impact on jet breakup. Yang and Turan (2017) have simulated liquid jet breakup using a Gerris solver. They have studied the effect of forced perturbation on low-speed jet breakup. Interestingly, in all papers using the Gerris solver, the jet inlet is given a flat velocity profile and sinusoidal temporal perturbation is added to initiate the breakup. The authors have also mentioned that to trigger jet breakup perturbation is required. This could be attributed to neglecting the velocity profile which triggers instability (Sinha, 2019; Yoon and Heister, 2004). Yoon and Heister (2004) have studied liquid primary atomization of an axisymmetric liquid jet using the boundary element method. They attribute boundary layer instability at the orifice exit to cause instability in the jet surface which is being modelled as a vortex ring. They have compared their results for various operating conditions with the experimental images by Hoyt and Taylor (1977). They claim to obtain similar morphological features on jet fragmentation. In a similar study Park and Heister (2006) have attributed the boundary layer thickness to affect the vorticity and location of vortex rings they model. Using the same numerical framework, Yoon (2005) investigated droplet generation from primary atomization of an axisymmetric liquid jet. It is observed that log-normal distribution gives the best fit with the obtained droplet sizes.

Gong et al. (2016) systematically studied the effect of injector geometry details on the formation of surface waves on straight liquid jets under atmospheric ambient. Injector geometry is understood to affect the boundary layer growth inside the injector tube (Sinha, 2023; Prakash et al., 2018), and impact the jet breakup. This observation is further explored with a theoretical model derivation incorporating injector geometry effects for a liquid jet in crossflow configuration (Sinha, 2019). This model matches well with the experimentally observed wavelength on liquid jets in crossflow.

Mehravaran (2013) has examined the impact of artificial perturbation in liquid jet breakup. He has concluded that the perturbation leads to enhanced secondary breakup of droplets and the average droplet size is reduced drastically as compared to an unforced jet. In recent papers using Basilisk (Balaji, 2021, Kumar, 2023), jet breakup is observed without artificial perturbation, when a parabolic velocity profile is used. Some recent works consider injector geometry effects to be significant and have incorporated this in their computational modeling. Jiao et al. (2017) carried out DNS of liquid jet breakup. To accurately capture the jet inlet turbulent characteristics, they have separately simulated single-phase flow inside the nozzle tube. Data from tube flow is used to define inlet conditions for the VOF simulations of atomizing jet, where the nozzle geometry is excluded from the computational domain. Agarwal and Trujillo (2020) have used X-ray tomographic data of nozzle internal geometry (ECN- Spray-A Nozzle) in their study. They have carried out a VOF study using realistic approximations of the nozzle's internal geometry. They observed that the internal surface features of the nozzle give rise to non-axial velocity components which aid early breakup. One case is also considered where the nozzle geometry effects are ignored and a uniform inlet velocity profile is considered for the jet. In that case, the jet is found to be most stable. This paper focuses on the local disturbance caused by nozzle internal surface features. However, the impact of the velocity profile is overlooked. Ghiji et al. (2016) have undertaken both experimental and numerical studies of a diesel jet injected in a high-pressure ambient. They have captured instantaneous jet images experimentally which matches reasonably well with computational results. They have included the injector geometry in their computational model and have simulated the injector tube to obtain a realistic velocity profile for the liquid jet.

From the preceding discussion, it can be summarized that there are two approaches available in the literature. A simplistic approach is to inject liquid as a plug flow, or flat/ uniform velocity profile (Pairetti et al., 2020; Zhang et al., 2020). Another approach could be to simulate flow inside the atomizer and use this velocity profile as an input to jet simulations (Agarwal and Trujillo, 2020; Jiao et al., 2017). The first approach essentially ignores the jet velocity profile effects and would not be able to capture surface instability accurately. The second approach goes to the other extreme where X-ray tomography is required to accurately capture the nozzle's internal surface. That might be more accurate but not practically feasible. In this paper, we present a third approach where a standard mathematical function is used as the jet velocity profile boundary condition and its impact on jet breakup is accessed. For realistic injector dimensions and flow rates, the liquid is expected to fall under the laminar regime. It is also believed that velocity redistribution will be more severe in laminar, as turbulent velocity profile is flatter, or more uniform. This paper presents a detailed analysis of jet breakup behavior for various velocity profiles. For this study, we have used a parabolic velocity profile for laminar flow and, a power-law velocity profile for turbulent flow. A flat velocity profile is also simulated for comparison.

## Modeling Details

### Governing Equations

The present study uses the Basilisk platform for all the simulations (Popinet, 2020). Basilisk is based on the Momentum-Conserving volume of Fluid (MCVOF) solver. The governing equations can be expressed as:

$$\nabla \cdot \vec{u} = 0 \qquad (1)$$

$$\frac{\partial}{\partial t}(\rho \vec{u}) + \nabla \cdot (\rho \vec{u} \vec{u}) = -\nabla p + \nabla \cdot (2\mu D) + f_\sigma \qquad (2)$$

Where $\vec{u}$ denotes velocity, $p$ stands for pressure, $\rho$ is for the fluid density, and $\mu$ is for fluid viscosity. Tensor $D$ can be expressed as:

$$D = \frac{1}{2}[\nabla \vec{u} + (\nabla \vec{u})^T] \qquad (3)$$

$f_\sigma$ is for the surface tension force:

$$f_\sigma = \sigma \kappa \vec{n_s} \delta_s \qquad (4)$$

$\sigma$ is the surface tension coefficient, $\kappa$ denotes the curvature, $\vec{n_s}$ normal and $\delta_s$ stands for the Dirac function. One fluid formation is used, where color function $c$ takes the value of 1 in the liquid phase and 0 in the gaseous phase. The transport equation for $c$ can be given as:

$$\frac{\partial c}{\partial t} + \nabla \cdot (c \vec{u}) = c \nabla \cdot (\vec{u}) \qquad (5)$$

For a more detailed description of the solver and numerical method, the reader is referred to (Popinet, 2020; Zhang et al., 2020; Pairetti et al., 2020).

**Computational Domain and Boundary Conditions**

The computational domain is shown in Fig. 1. It is a cube of side 15 D, where D is the injector diameter. The left boundary is a wall with a hole at its center. The hole is the fuel inlet, and the liquid entry conditions are specified on this boundary. The right boundary is the pressure outlet. All other boundaries are designated as no-shear walls. All the cases are run for the same operating conditions, except for the velocity profile. Fluid properties and relevant parameters are given in Table 1.

| Property/ Parameter | Value |
|---|---|
| Jet diameter - D (mm) | 0.1 |
| Average jet velocity - $U_{avg}$ (m/s) | 100 |
| Liquid density (kg/m$^3$) | 848 |
| Gas density (kg/m$^3$) | 34.5 |
| Density Ratio | 24.58 |
| Liquid viscosity (Pa. s) | 2.87 x 10^ |
| Gas viscosity (Pa. s) | 1.97 x 10$^{-5}$ |
| Surface tension coefficient (N/m) | 0.03 |
| Jet Reynolds number | 2955 |
| Aerodynamic Weber number | 28267 |

**Table 1. Fluid Properties and Operating Conditions**

# Results and Discussion

As discussed in the previous section, three velocity profiles are used- parabolic, power-law, and flat. All these velocity profiles are depicted in Fig. 2. Average jet velocity is maintained the same for all the cases. Further, the time scale is non-dimensionalized using jet average velocity ($U_{avg}$) and jet diameter ($D$):

$$t^* = \frac{D}{U_{avg}}  \qquad (6)$$

Jet structures resulting from different velocity profiles are shown in Fig. 3. All the images are taken at t/t* = 9. The c

As evident from Fig. 2, even for the same average velocity, the peak velocity for a parabolic profile is two times that of the flat profile. Peak velocity is highest for the parabolic case and lowest for the flat velocity profile. The difference in peak velocity values results in the penetration of the parabolic jet being the highest and the flat jet being the lowest. Power-law jet falls in between these two extremes. Now, velocity profiles also affect the jet tip shape. The tip of the parabolic case is small, aerodynamically rounded, and faces lower air resistance which aids in jet penetration. The flat profile has a flat tip, which further distorts and becomes bulbous, making it difficult to penetrate. The tip of the power-law jet is better shaped than the flat jet but is less aerodynamic than the parabolic case.

Figure 4 shows the droplet size distribution for all three cases. The number of droplets for each diameter size is shown. Droplet numbers are normalized using the highest number (for any diameter) to obtain a Probability Density Function (PDF). The drop-size distribution appears similar for all the cases, while the parabolic case looks slightly broader than the other cases. Further, the Rosin-Rammler distribution is assumed, and corresponding parameters are derived from curve-fitting. The curve fit appears to be within a reasonable tolerance. Rosin-Rammler, also known as the Weibull distribution, is a widely used distribution for droplets in sprays (Lefebvre and McDonell, 2017). It can be expressed as:

$$f(d) = a \cdot b \cdot d^{b-1} \cdot \exp(-a \cdot d^b) \qquad (7)$$

Where *a* and *b* are constants, and *d* is the diameter bin size. Their values for various cases are listed in Table 1.

| Case | a | b |
|------|-------|-------|
| Parabolic | 0.594 | 2.672 |
| Power-law | 0.647 | 2.635 |
| Flat | 0.602 | 2.463 |

**Table 2. Rosin-Rammler constants for various cases**

To gain further insight into the breakup behavior and its dependence on the jet velocity profile, atomized volume is examined as a percentage of total injected volume. Fig. 5(a) shows how the percentage of atomized volume increases with time. Since the average velocity is the same for all the cases, the total injected volume at any time instant will be the same. As evident, a flat profile shows the maximum atomized volume at all times, while parabolic and power-law cases have lower and comparable atomized volumes. As observed in jet structural images, the flat case has a bulbous and large head tip which is responsible for producing numerous small droplets and consequently a larger percentage of atomized volume. Almost all of the droplets observed in the initial phase of breakup are produced by the jet head tip. This aspect will be discussed in detail in later sections. Further, distortion in droplets is assessed by plotting the atomized volume fraction with sphericity ($\psi$) in Fig. 5(b). Sphericity can be defined as (Pairetti et al., 2020):

$$\psi = \frac{D_{32}}{D_{30}} \tag{8}$$

Where $D_{32}$ and $D_{30}$ are equivalent droplet diameters defined as:

$$D_{32} = 6\frac{V_d}{A_d} \tag{9}$$

$$D_{30} = \left(6\frac{V_d}{\pi}\right)^{(1/3)} \tag{10}$$

and $V_d$ and $A_d$ are droplet volume and surface area, respectively. A perfect sphere will have $\psi=1$ and a distorted droplet will have $\psi < 1$. Smaller values correspond to larger distortion. As observed in Fig. 5(b), droplets from the power-law profile are slightly more distorted than in other cases. Parabolic and flat profiles have similar levels of droplet distortion. However, the difference between cases is not appreciable. In general, droplets appear to be near-spherical in shape. This could partly be attributed to their small size. Since the liquid jet is of diameter 0.1 mm, the droplets produced are also very small.

To understand surface instability and jet breakup, axial velocity contours are plotted in the mid-plane section for all cases in Figure 6. The color represents the axial velocity magnitude. The color bar is selected to vary from 0 to 80% of the maximum velocity. The same color bar is followed in all Figures. As evident, the flat profile has uniform velocity from the jet center to the jet boundary and there is no boundary layer in the near injector exit region. The boundary layer grows gradually due to air resistance. In contrast, the parabolic and power-law jets have a visible boundary layer right from the injection point. To probe the temporal evolution of the jet, and boundary layer, jet mid-plane images are shown for different time instances. Figure 7 shows the evolution of a flat velocity profile jet. The jet tip growth is nicely captured in this image. The jet tip is observed to get larger with time. This could be attributed to the air resistance blocking the non-aerodynamic jet tip. Surface instability appears near the jet tip and grows with time. Also notice how the velocity values show a jump at the jet edge, which is typical for a flat velocity profile. Figure 7 depicts the temporal evolution of the power law jet. In contrast to the previous case, a distinct boundary layer is observed near the jet edges, which grows with time. The jet tip appears to be pointed initially but gradually changes to a round shape. Again, air resistance plays a role in making the tip rounder. Surface breakup is seen almost for the entire jet length. Finally, the evolution of the parabolic case is depicted in Figure 9. As discussed, parabolic cases show maximum penetration. The velocity boundary layer at the jet surface is distinctly visible. It is more prominent than the previous cases.

Figure 10 shows the velocity profiles for all cases plotted at different axial locations. Velocity profiles are shown for 1D, 5D, and 10D for parabolic cases in Figure 10(a). At the 1D location, the jet has just emerged from the injector and maintains a parabolic profile, with the peak velocity closely matching the jet injection boundary condition. Although near the edges, instability is evident, even for 1D. This surface instability is further amplified at 5D. The velocity at the center decreases gradually as the jet slows down while penetrating against air resistance. It is also apparent that the jet is getting narrower, which is evident in the jet structure images discussed previously. The velocity profile for the power-law case is shown in Figure 10(b). Initially, it shows a power-law profile at 1D, which is slowed down at downstream locations of 5D and 7D. Jet velocity at the edge increases for 5D, which is also evident in the parabolic case. Velocity redistribution is believed to be one of the factors triggering instability on the jet surface (Sinha, 2019). Increasing velocity near the edge at the 5D location appears to support this hypothesis. The last case of the flat profile is shown in Figure 10(c). As clear from jet structure images, flat jet penetrates the least among all the cases investigated here. Hence, velocity profiles for only 1D and 5D locations are obtained. The jet shows a flat profile at 1D, with signatures of a boundary layer. The boundary layer is not visible at 5D due to ligament rupture and jet surface breakup near the edges. Further information can be obtained by comparing the derivatives of axial velocity concerning radial coordinates. The first and second derivatives of velocity are compared in Figure 11 for the same axial locations as shown in Figure 10. The first columns of Figure 11 show the first derivative ($\partial u/\partial r$), and the second column shows the second derivative ($\partial^2 u/\partial r^2$). The value of ($\partial u/\partial r$) is always negative for the parabolic profile, as shown in Fig. 11(a). This is expected, as an ideal parabolic function has zero slope at the center and a negative slope at the edge. Similar behavior is also evident in the power law profile plotted in Fig. 11(b). Whereas, the flat profile has zero slope for most of the jet radius. Only at the edge does it slightly decrease. This can be attributed to the nature of the flat profile. For an ideal flat profile, the slope should be zero everywhere, but due to surface instability, there is a small decrease near the edges.

The flat profile is unphysical as it violates the no-slip boundary condition. Hence, for further discussion, only parabolic and power-law jets will be considered. Figure 12 presents the jet structure colored by the axial velocity magnitude. Fig. 12(a) shows the parabolic case while the power-law case is displayed in Fig. 12(b). The power-law jet is zoomed in to highlight the rich jet structure. The parabolic case demonstrates the process of velocity redistribution which is understood to be the prime cause of surface instability (Sinha, 2019). The jet surface appears to be at almost zero velocity till some distance from the nozzle. Subsequently, due to momentum exchange with the core liquid, the surface velocity gradually increases. This is evident in the change of color of the jet surface along the axial direction. Similar behavior is also expected in the power-law case. However, in comparison to the parabolic jet, the slope of the power-law jet is much steeper near the jet edge. This results in velocity redistribution happening much more rapidly. Hence, the jet surface does not show near zero velocity near the injector exit.

The parabolic jet is explored in magnified images in Figure 13. Different sections of the jet are shown in various sub-images. Fig. 13(a) shows the initial instability and two-dimensional wave growth which is gradually giving rise to three-dimensional waves. The growth of these three-dimensional waves is captured in Fig. 13(b). The average wavelength of the waves is around 0.3 D. Further the jet tip and multiple pinch-off small structures are visible in Fig. 13(c). Although Fig. 13(a) and (c) show a large number of droplets, droplets are generally produced at the tip region. The droplets have lower velocities than the bulk jet. Hence, the droplets visible in Fig. 13(a) and (b) are pinched off from the near tip region earlier when the jet tip was passing through this region. The jet tip region is further zoomed in in Fig. 14. This image shows the multitude of smaller structures formed near the jet tip. The sheet thinning is observed, forming ligaments and smaller droplets.

Fig. 15 shows different sections of the power-law jet. As evident, the surface velocity is higher than that of a parabolic jet, which is expected for these velocity profile functions. The two-dimensional wave growth and transformation to three-dimensional waves are also observed in this case (cf. Fig. 15(a)). The wave has an average wavelength of 0.26 D. Further, Fig. 15(b) shows the jet tip. The jet tip is much flatter as compared to the narrow and pointed parabolic tip. The jet tip is further shown in zoomed-in images in Fig. 16. Sheet thinning is observed to be the main breakup mechanism at the tip surface, similar to a parabolic jet. Although the ligaments in this case are slightly larger than the parabolic jet. The degree of atomization is also lower as compared to the parabolic jet which was previously confirmed by droplet statistics (cf. Fig. 5(b)). The side view of both the jets showing the jet tip and surrounding droplets is shown in Fig. 16. The Parabolic jet is smaller than the power-law jet. The parabolic case also appears to have a higher number of droplets as compared to the power-law case which is already discussed in previous paragraphs.

## Conclusions

The paper presents a DNS study of liquid jet breakup using various jet velocity profiles. Previous studies have used flat or uniform velocity profiles which violates the no-slip condition imposed by the injector tube. Further, a parabolic velocity profile can be used to model laminar, and a power-law profile can be used to approximate turbulent regimes. All three cases are compared for jet structure, droplet size distribution, and degree of atomization. It is observed that the flat velocity profile produces the greatest degree of atomization, among these three cases. This can be attributed to the head vortex which breaks down into ligaments producing droplets. Power-law profile jet is observed to have the smallest degree of atomization. To gain an understanding of the jet structure, velocity distribution in jet mid-plane and its evolution with time is investigated. The velocity profile along the jet is also probed to obtain information on boundary layer stability. Further, morphological structures formed on jet surfaces are compared in detail for parabolic and power-law profiles. The jet surface is observed to show two-dimensional instability first and then transform to three-dimensional waves which subsequently produce ligaments and other surface features. It can be concluded that the jet surface instability and subsequent breakup are critically controlled by the jet inlet velocity profile which needs to be considered in any numerical simulation.

## Acknowledgments

The authors gratefully acknowledge the financial support provided by AR&DB for this project. The authors also acknowledge the support in the form of computational facilities from PARAM Shivay (IIT-BHU) for this work.

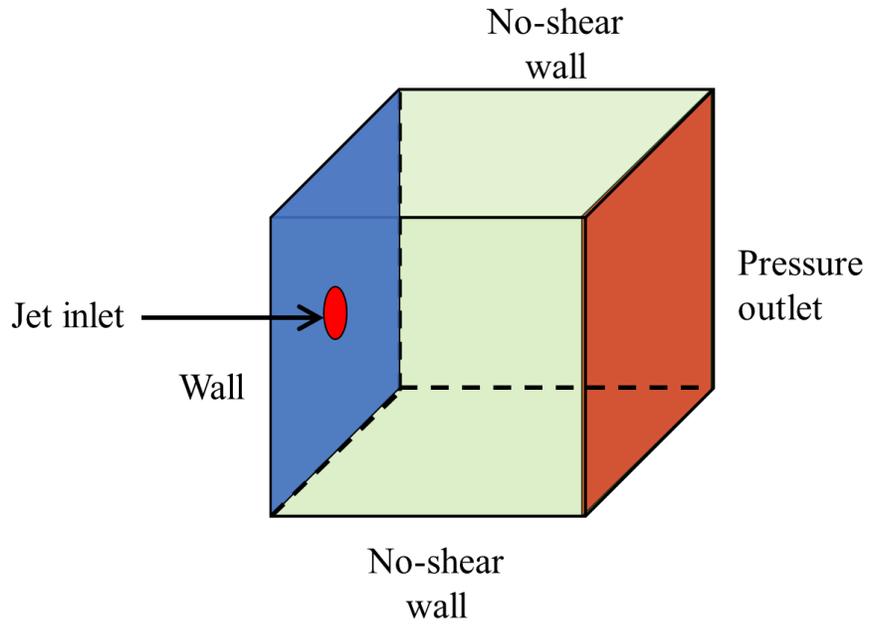

Figure. 1. Computational domain

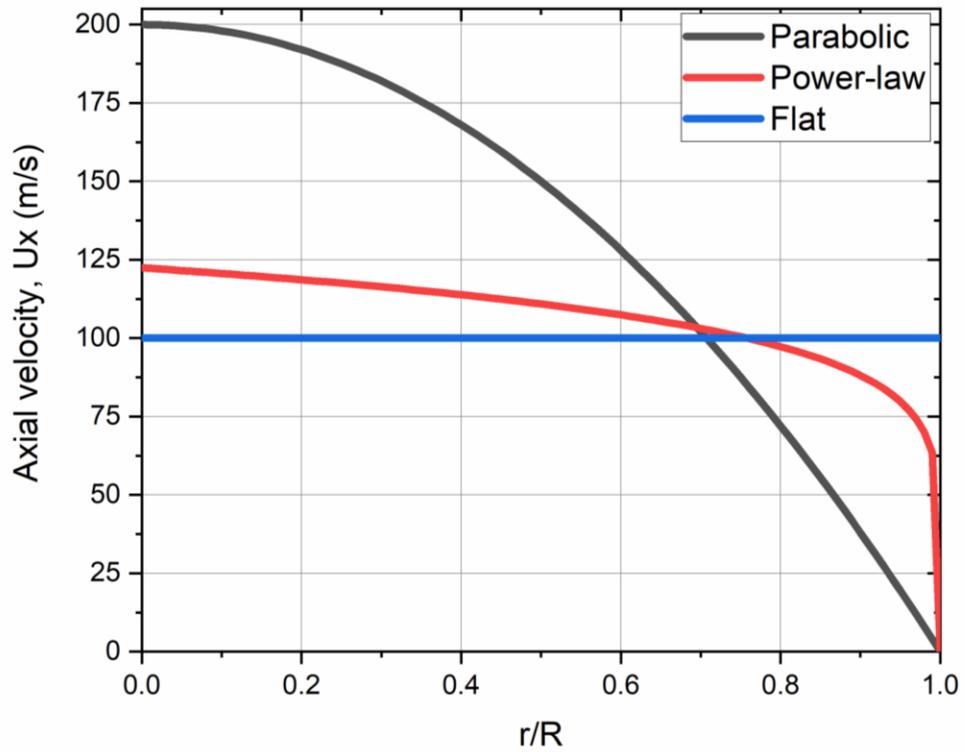

Figure. 2. Velocity profiles used for this study – parabolic, power-law and flat.

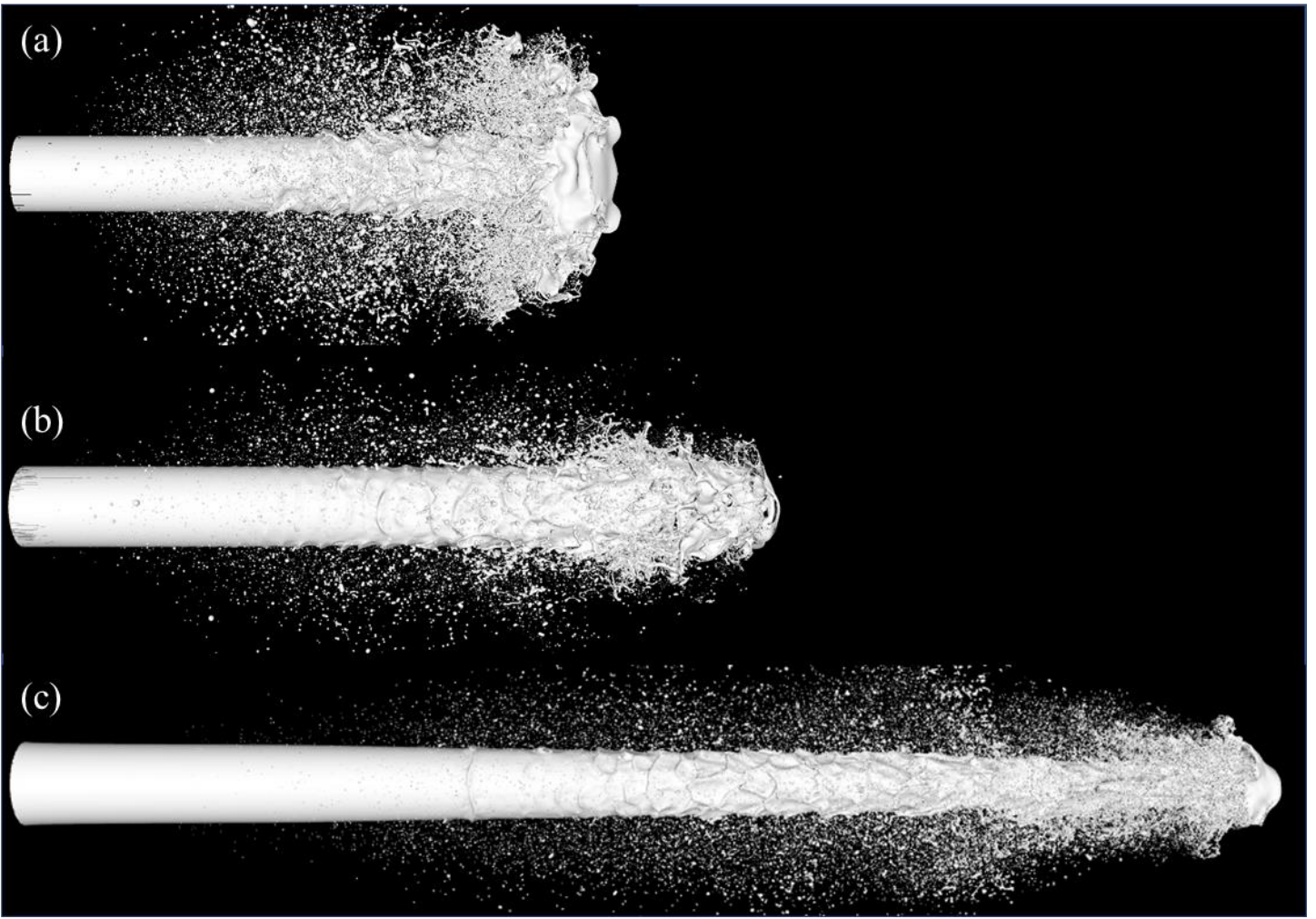

Figure 3. Jet structure for (a) flat (b) power-law, and (c) parabolic velocity profiles. All the images are for the same t* (t*=9).

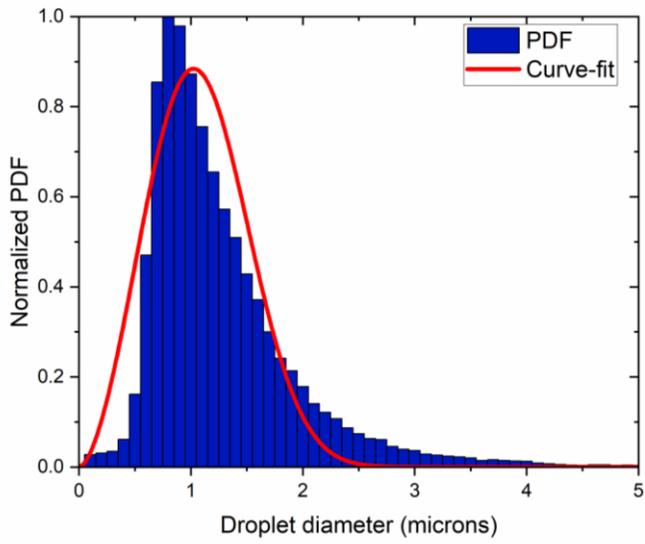
(a) Parabolic

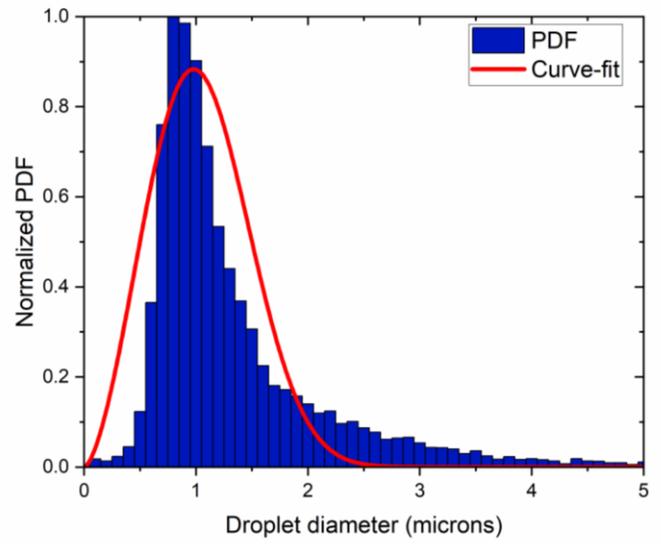
(b) Power law

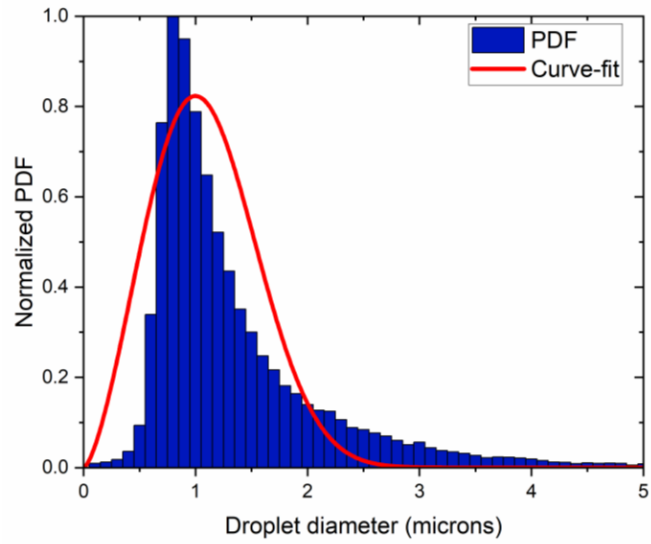
(c) Flat

Figure 4. Droplet size distributions plotted as Probably Density function (PDF) and curve-fits for (a) Parabolic, (b) Power law and (c) Flat velocity profiles.

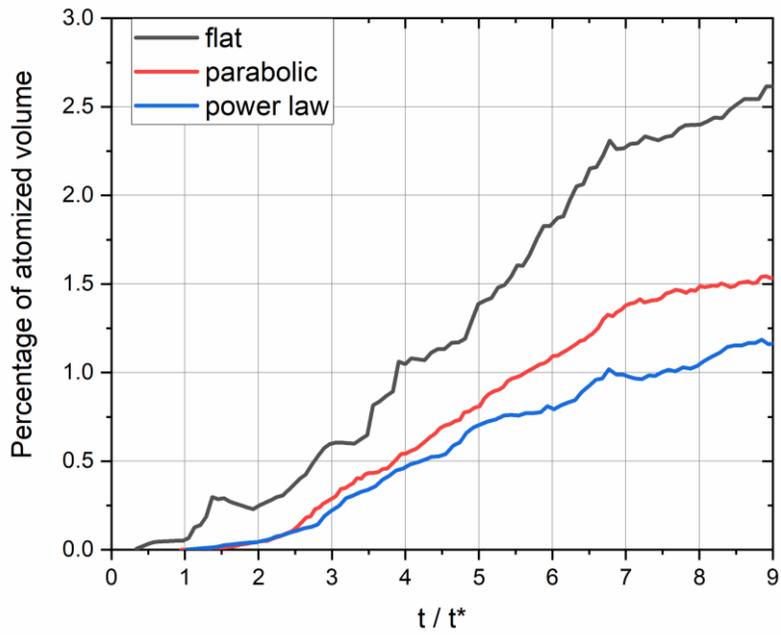

(a) Percentage of Atomized volume with t*

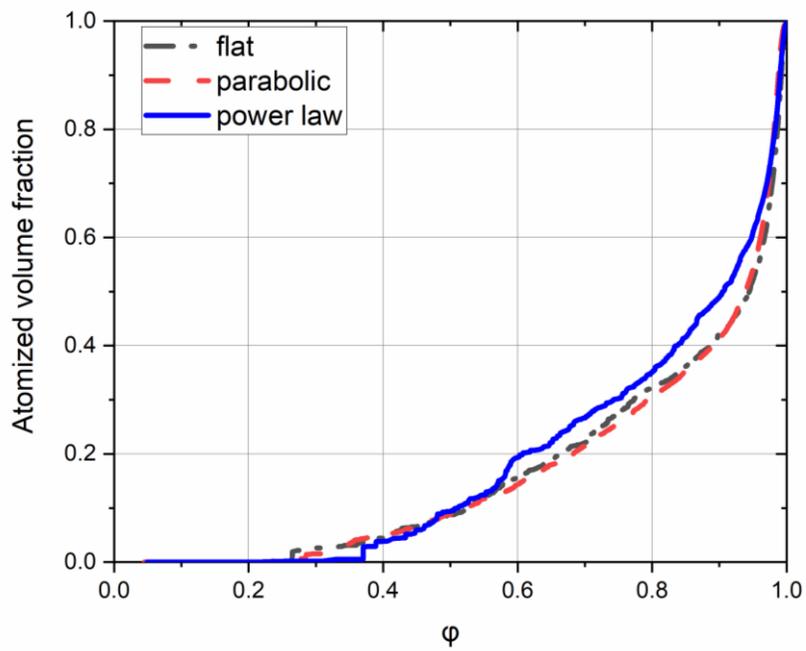

(b) Atomized volume fraction with distortion (φ)

Figure 5. Droplet statistics for various velocity profile jets.

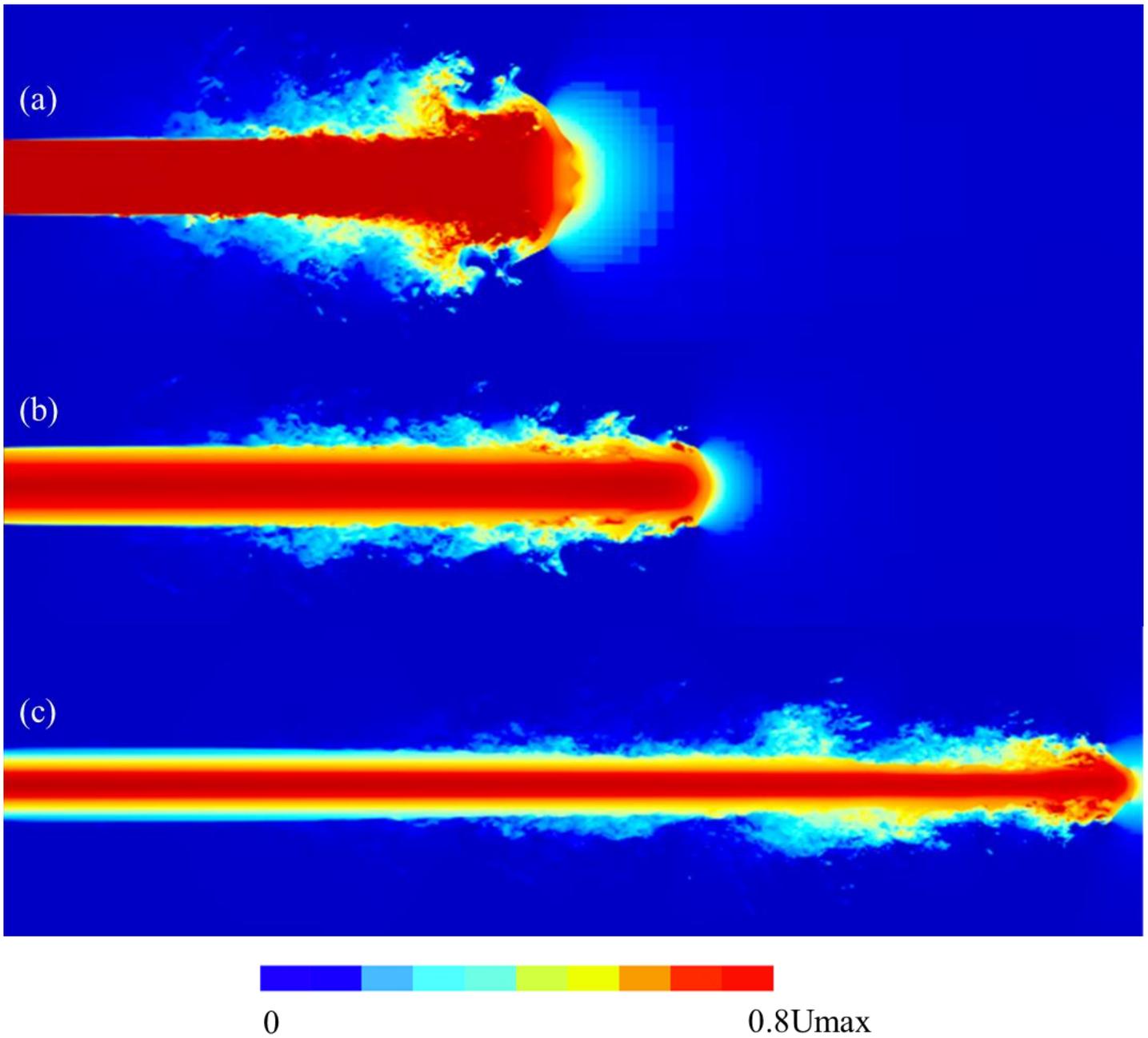

Figure 6. Axial velocity variation in the mid-plane for (a) Flat, (b) Power-law, and (c) Parabolic velocity profile jets. All the images are for the same t* (t*=9).

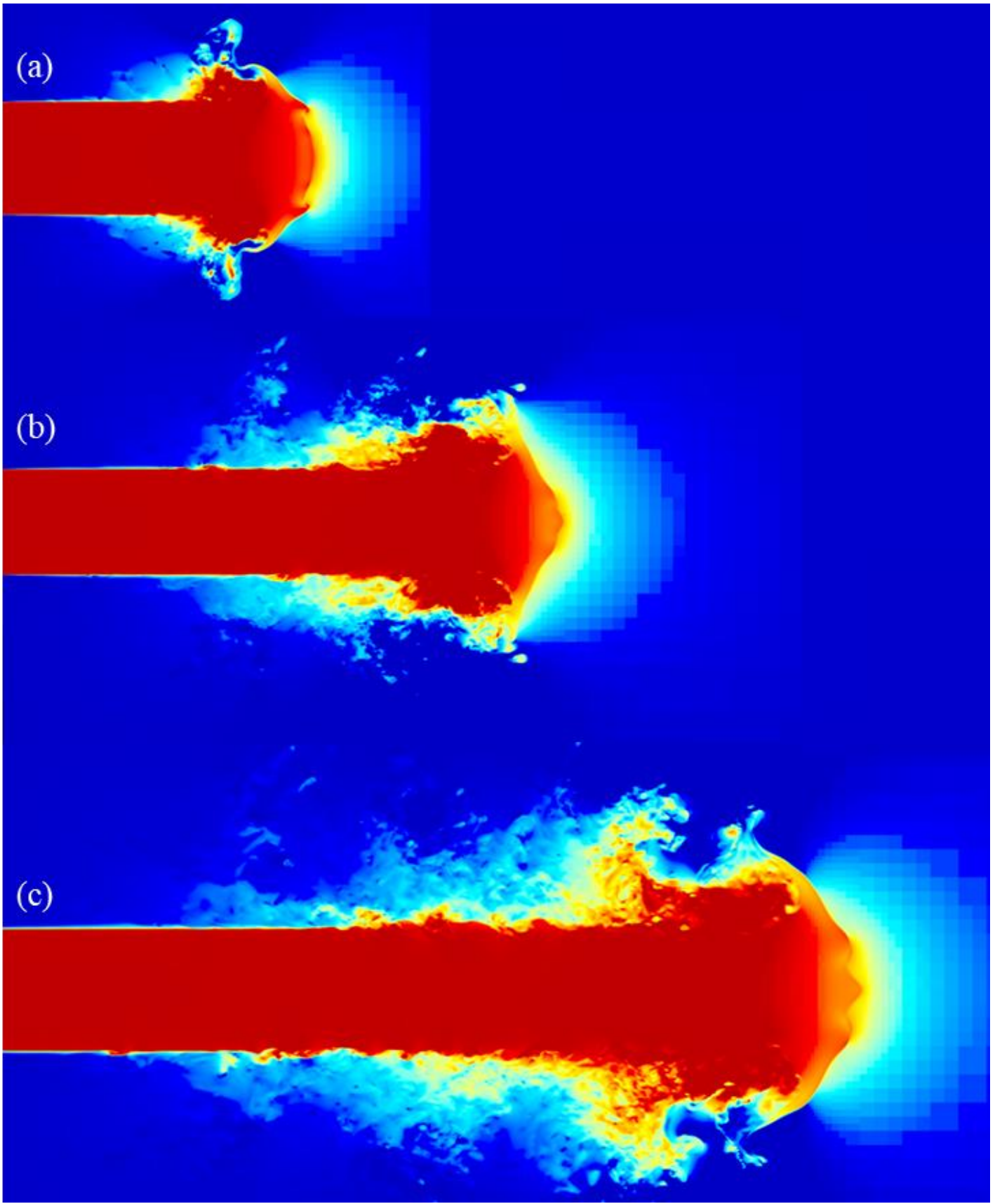

Figure 7. Evolution of flat velocity profile jet – axial velocity in the mid-plane section for (a) t*=3, (b) t*=6, and (c) t*=9.

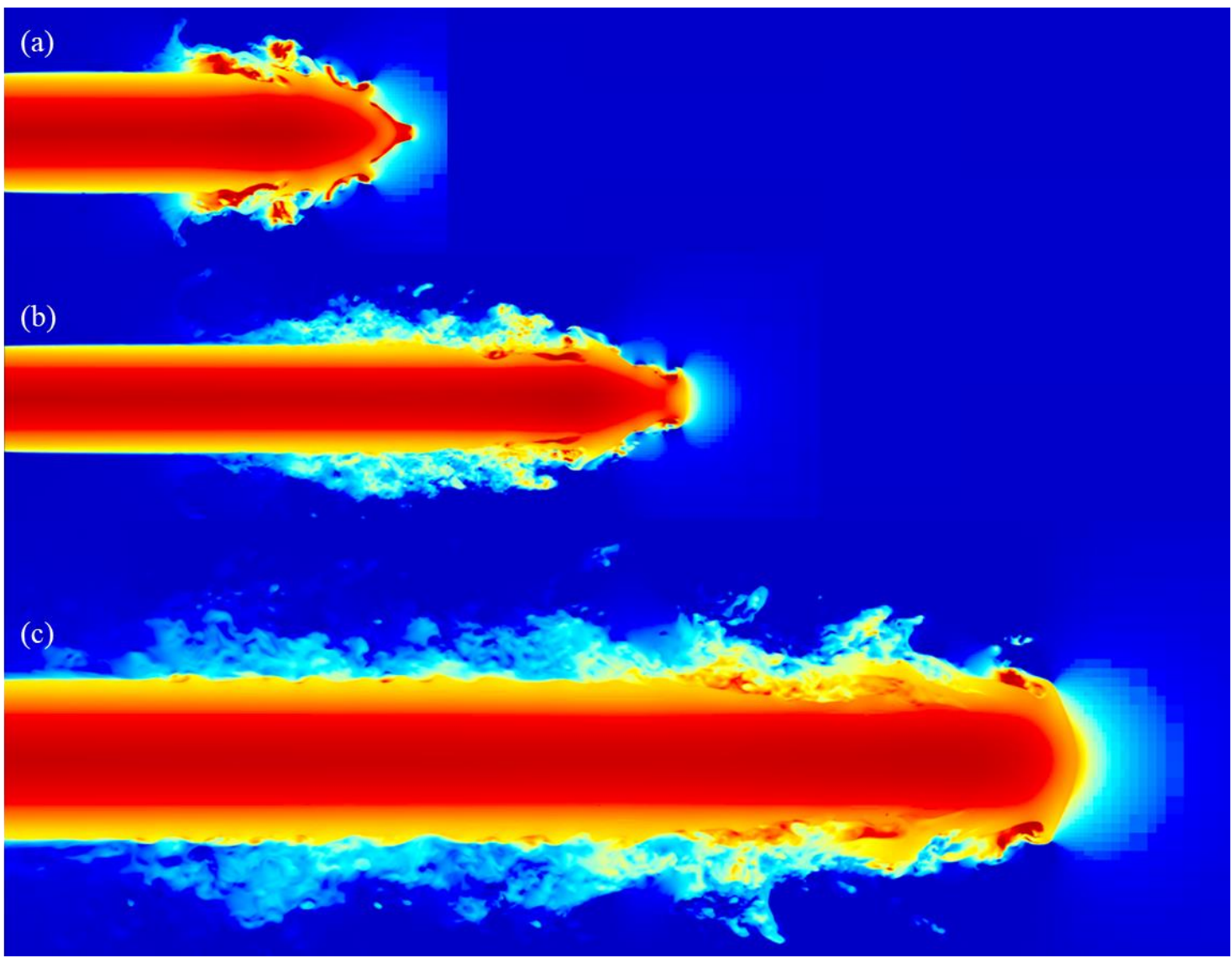

Figure 8. Evolution of power law velocity profile jet – axial velocity in the mid-plane section for (a) t*=3, (b) t*=6, and (c) t*=9.

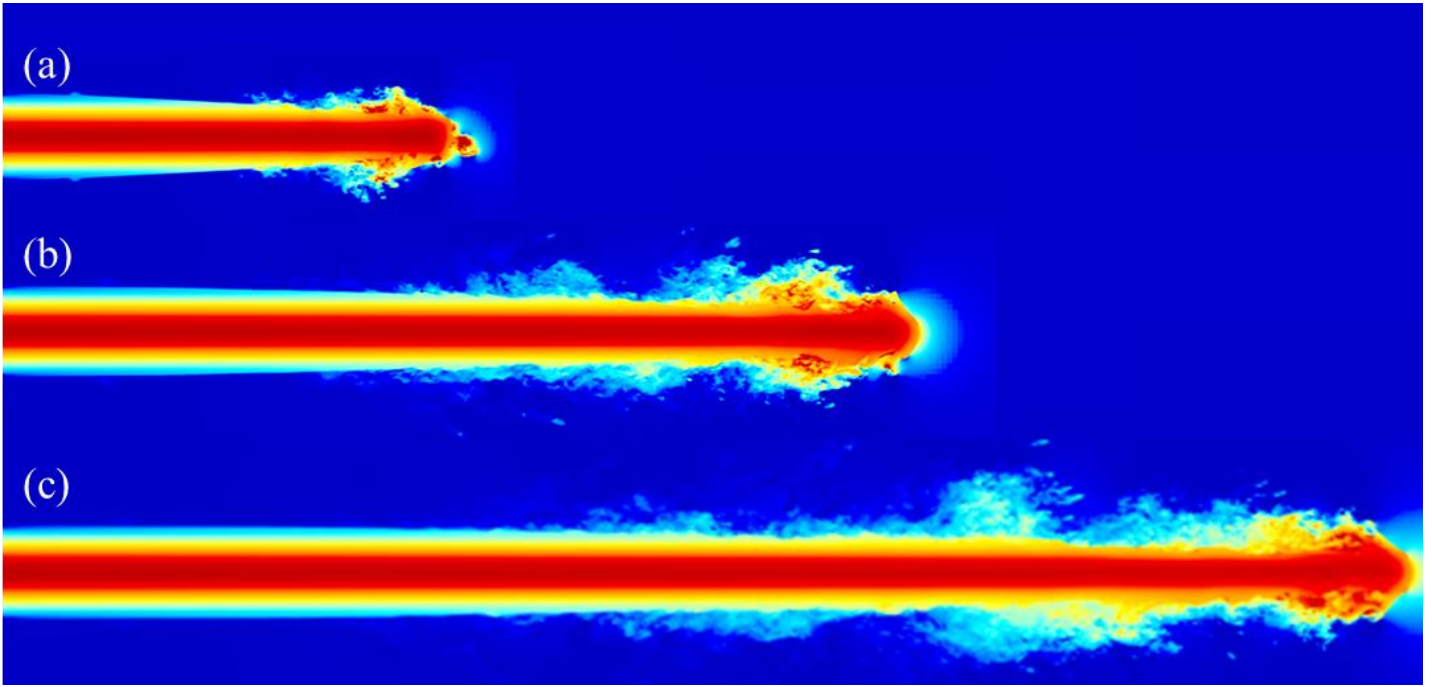

Figure 9. Evolution of parabolic velocity profile jet – axial velocity in the mid-plane section for (a) t*=3, (b) t*=6, and (c) t*=9.

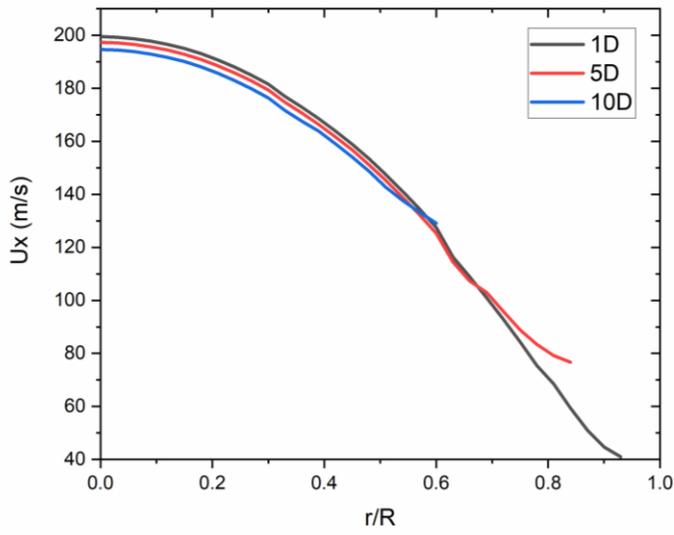
(a) Parabolic

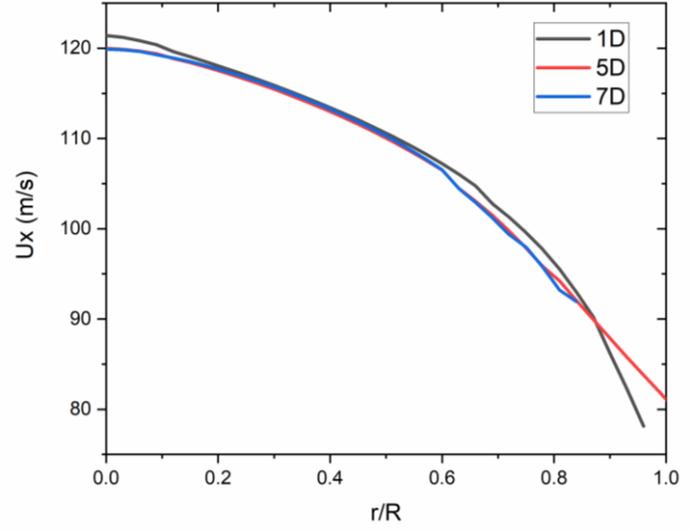
(b) Power law

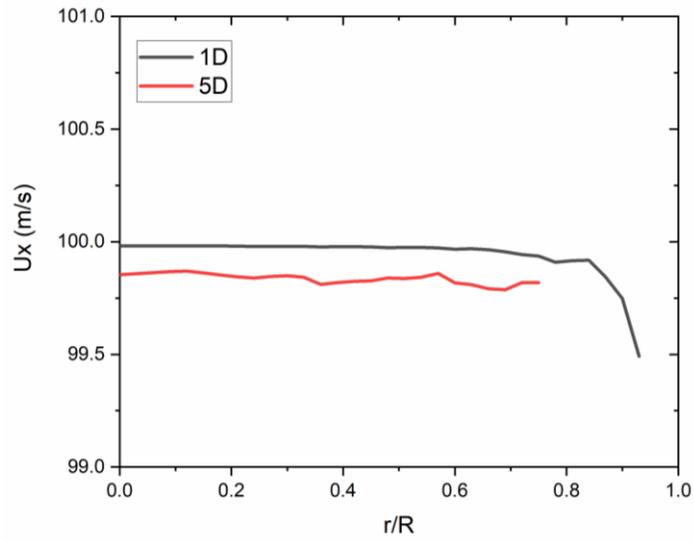
(c) Flat

Figure 10. Velocity profile obtained at various axial locations at t*=9, for (a) Parabolic, (b) Power-law, and (c) flat jets

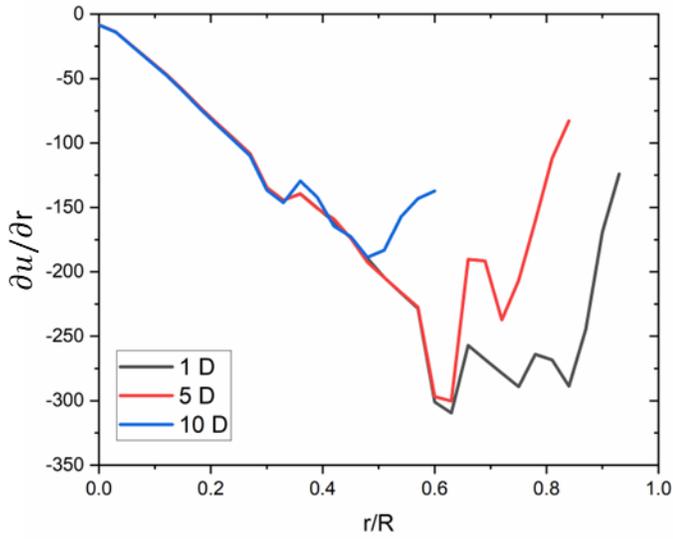
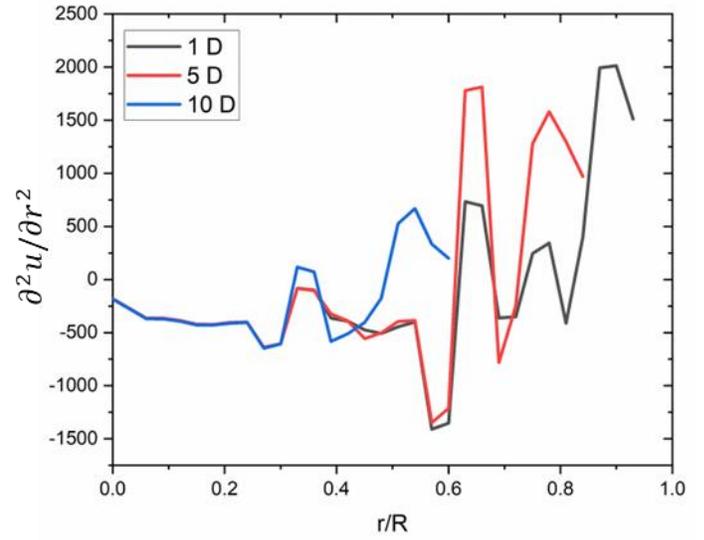

(a) First derivative of axial velocity (u), parabolic jet  (b) Second derivative of axial velocity(u), parabolic jet

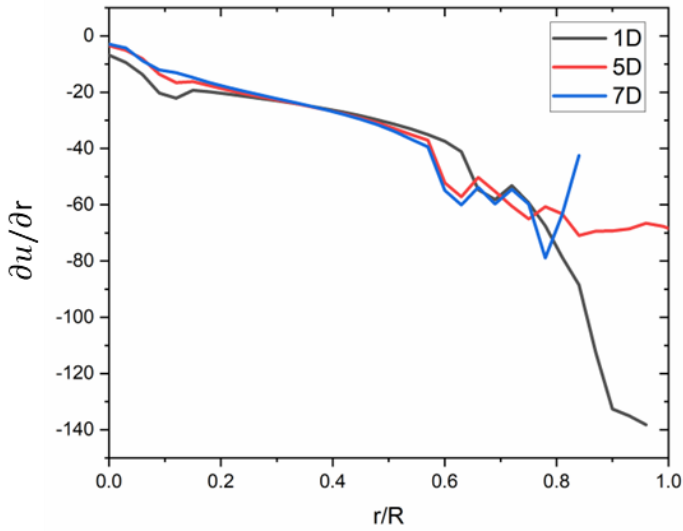
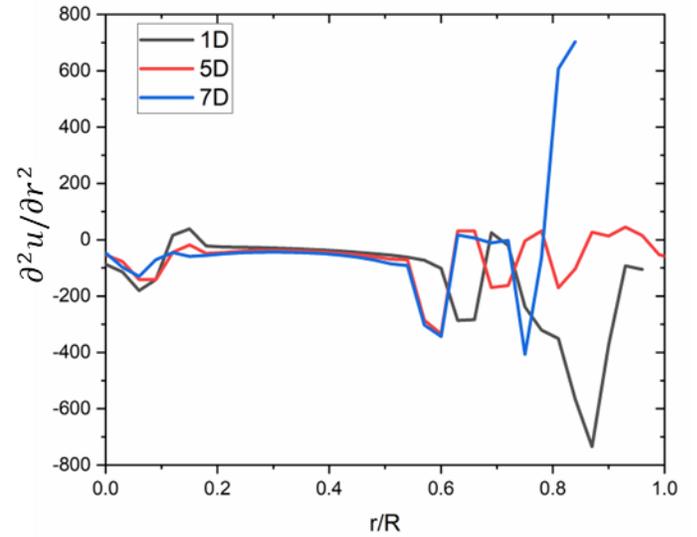

(c) First derivative of u, power law jet  (d) Second derivative of u, power law jet

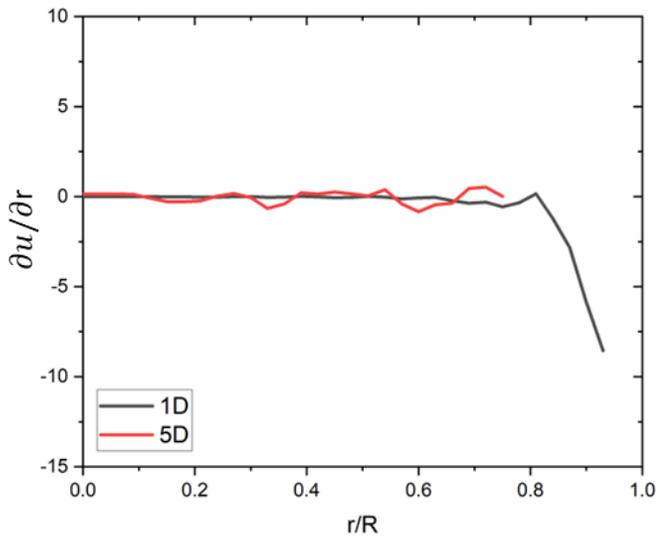
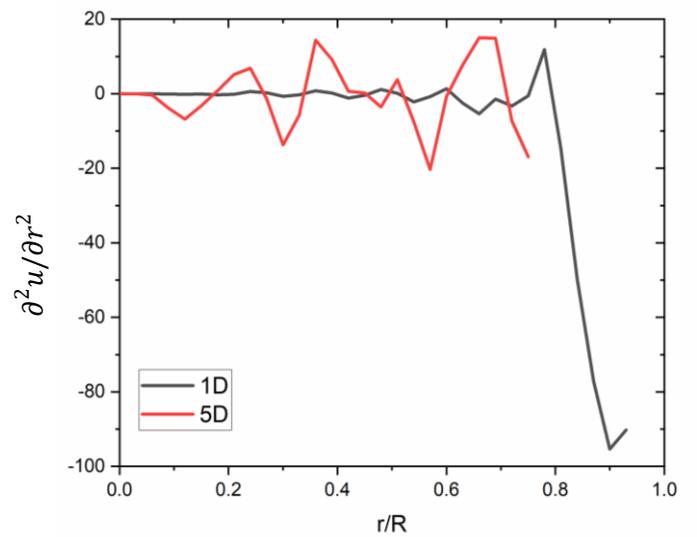

(e) First derivative of u, flat jet  (f) Second derivative of u, flat jet

Figure 11. First and second derivative of axial velocity obtained at various axial locations at t*=9

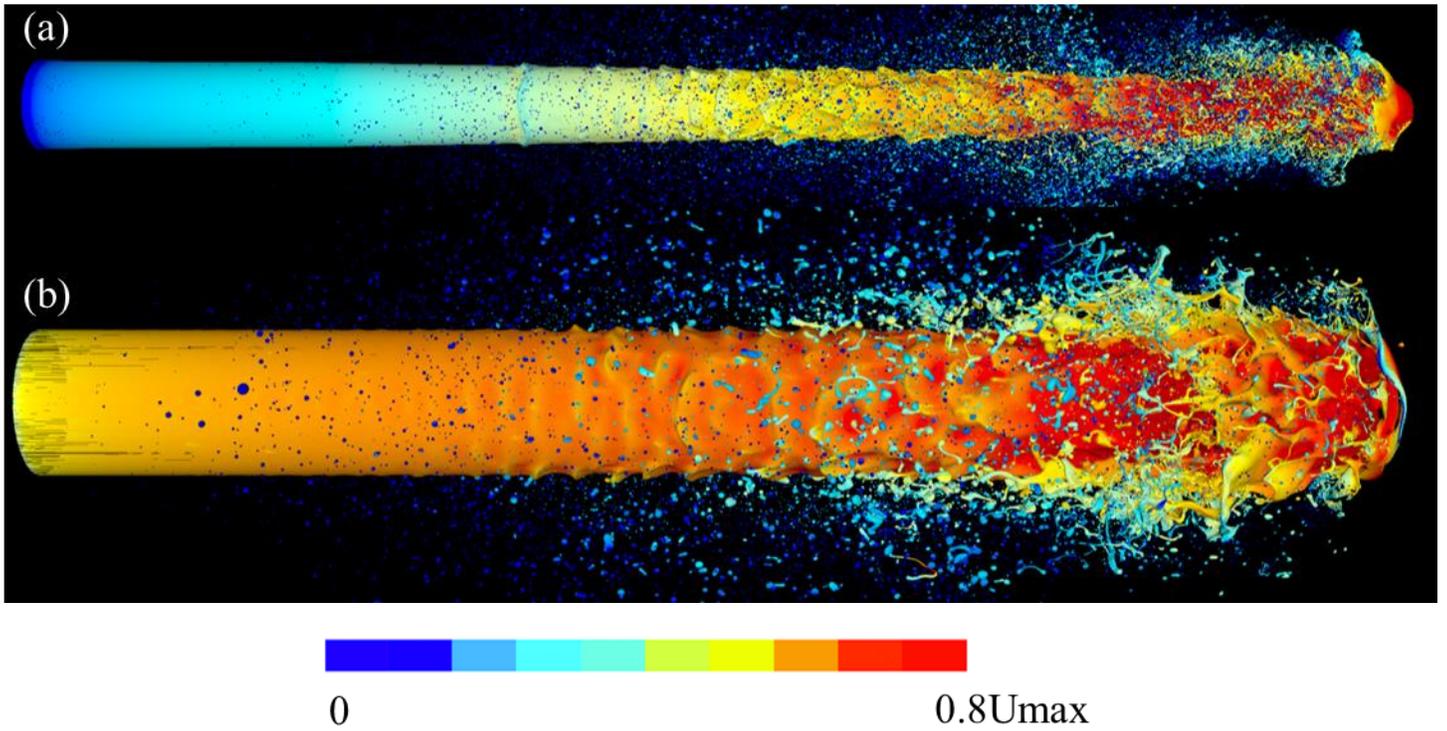

Figure 12. Structure of (a) parabolic, and (b) power-law jet at t*=9. Jet surface is colored by axial velocity (*U*). the color bar varies from 0 to 80% of Umax.

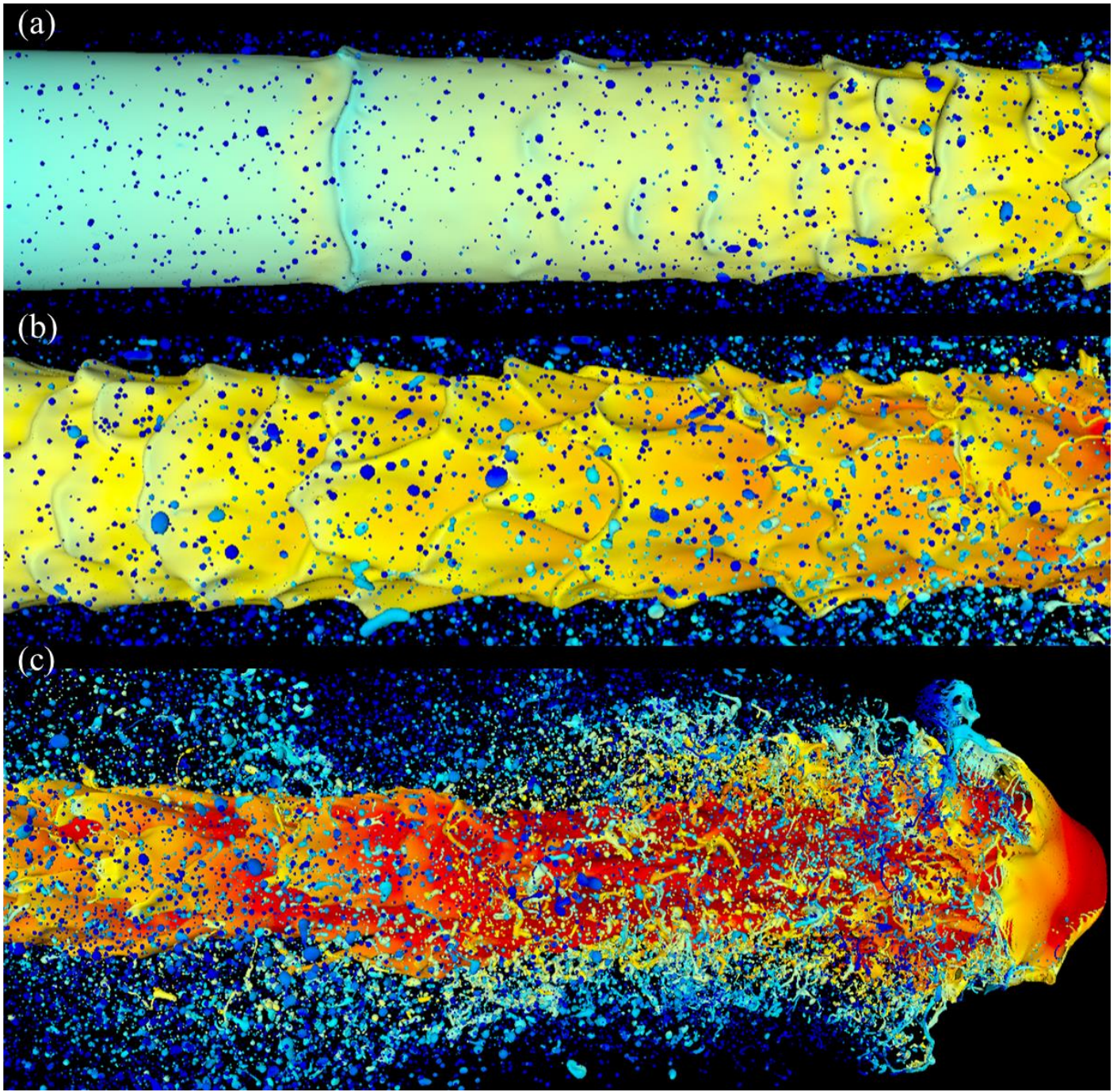

Figure 13. Structure of parabolic jet at t*=9. Jet surface is colored by axial velocity (*u*)

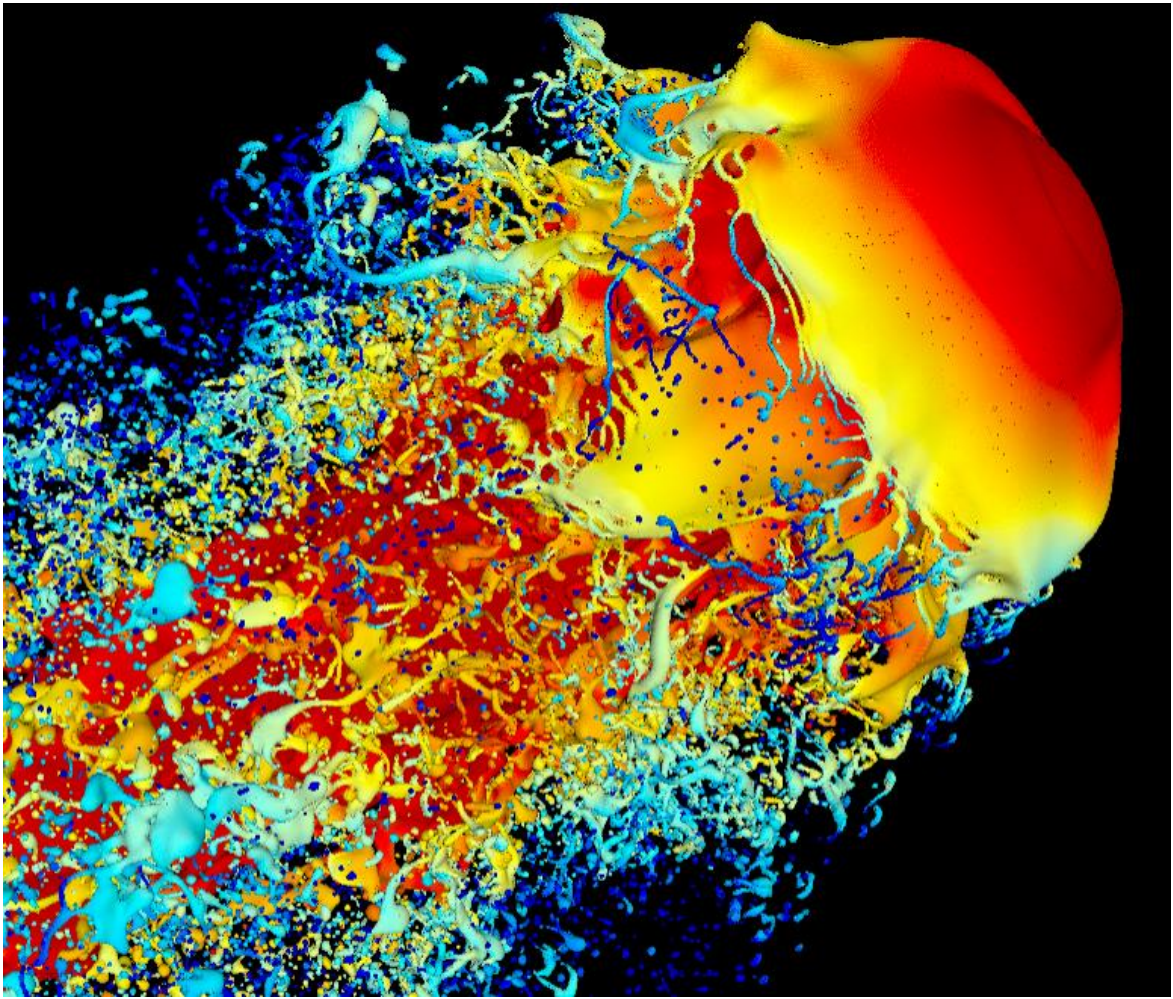

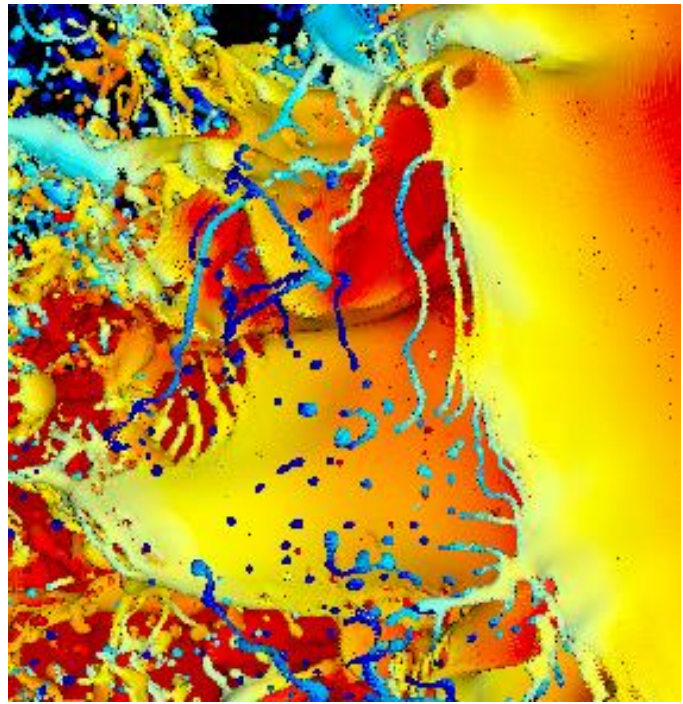

Figure 14. Structure of parabolic jet tip and sheet thinning at t*=9. Jet surface is colored by axial velocity (*u*)

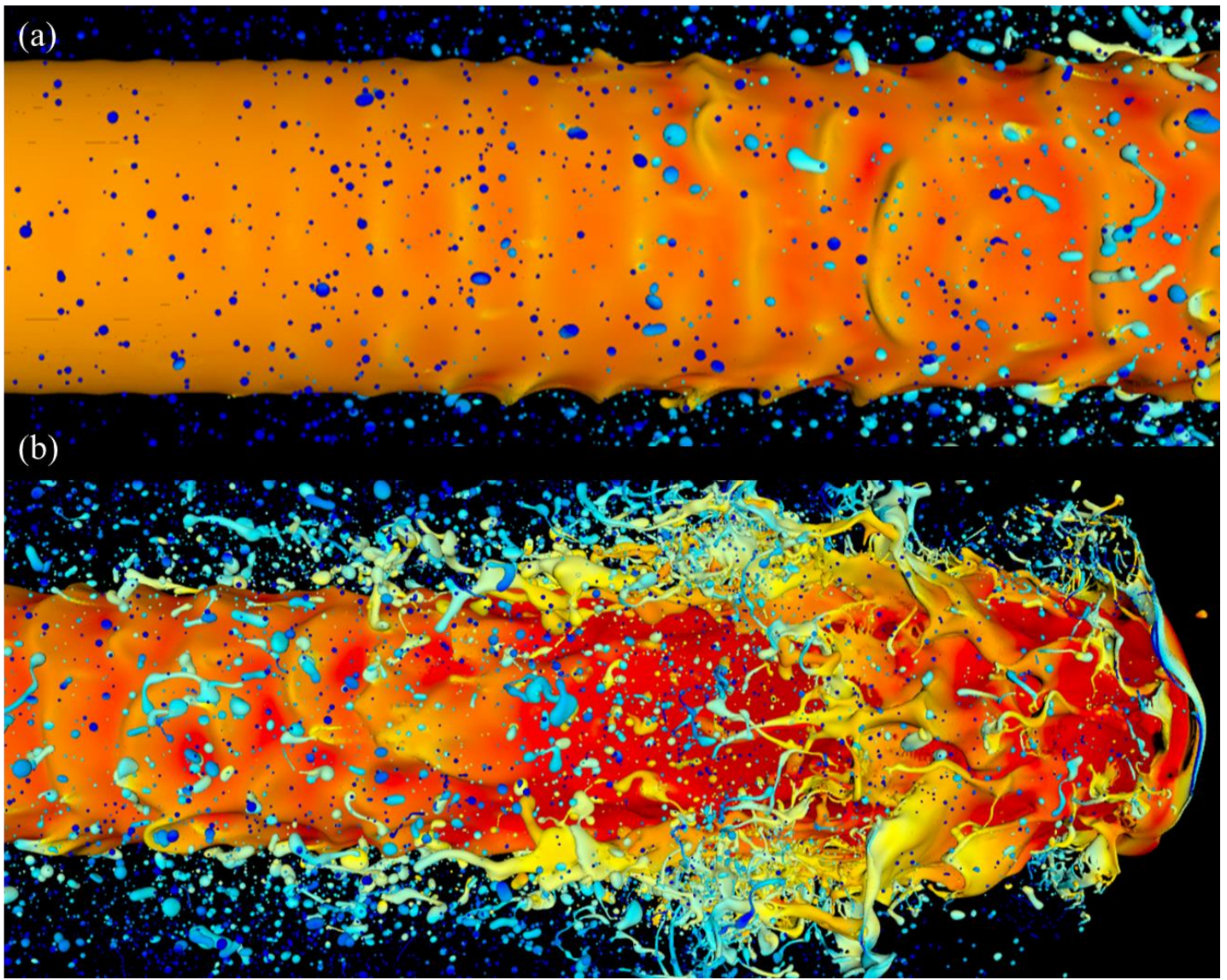

Figure 15. Structure of power-law jet at t*=9. Jet surface is colored by axial velocity (*u*)

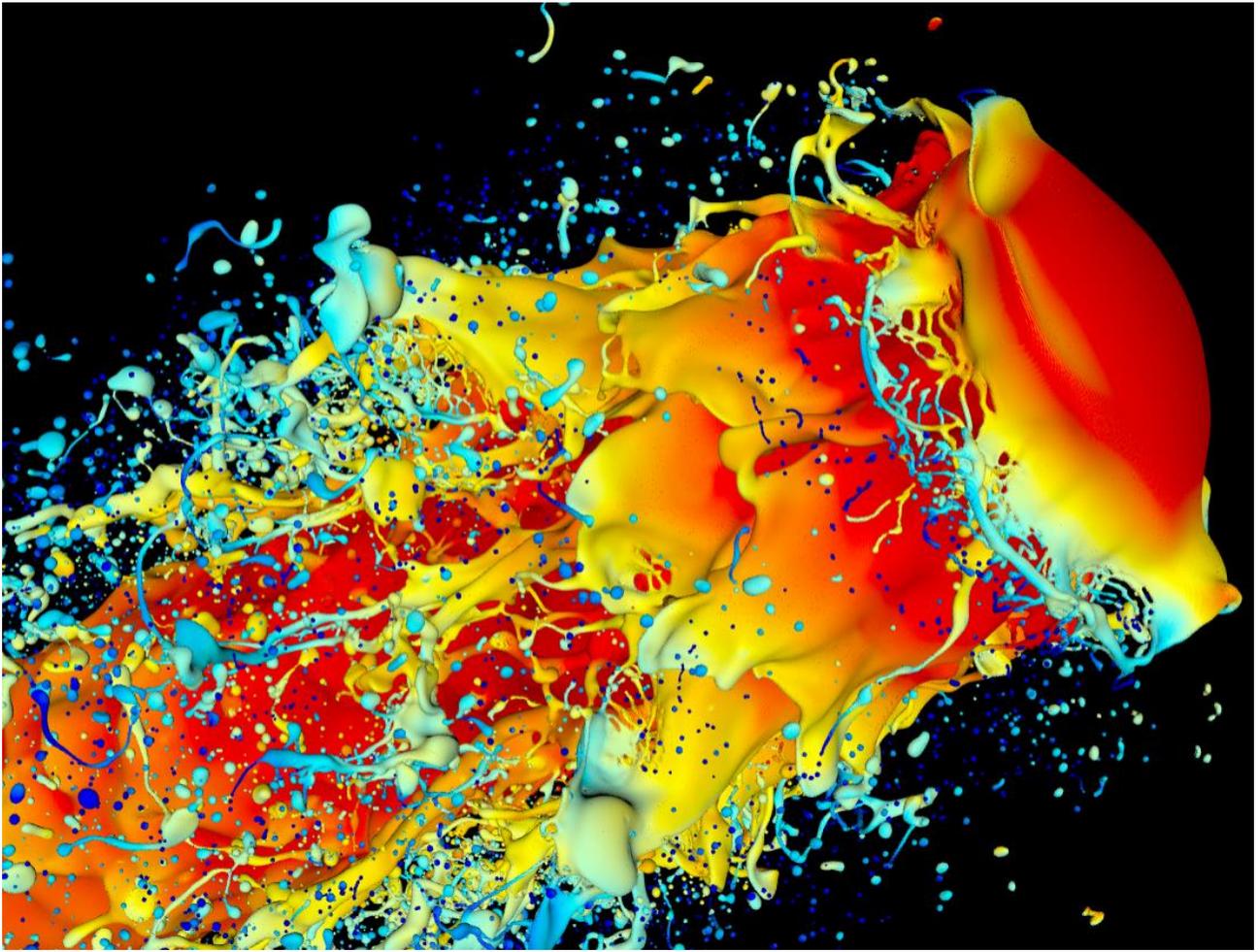
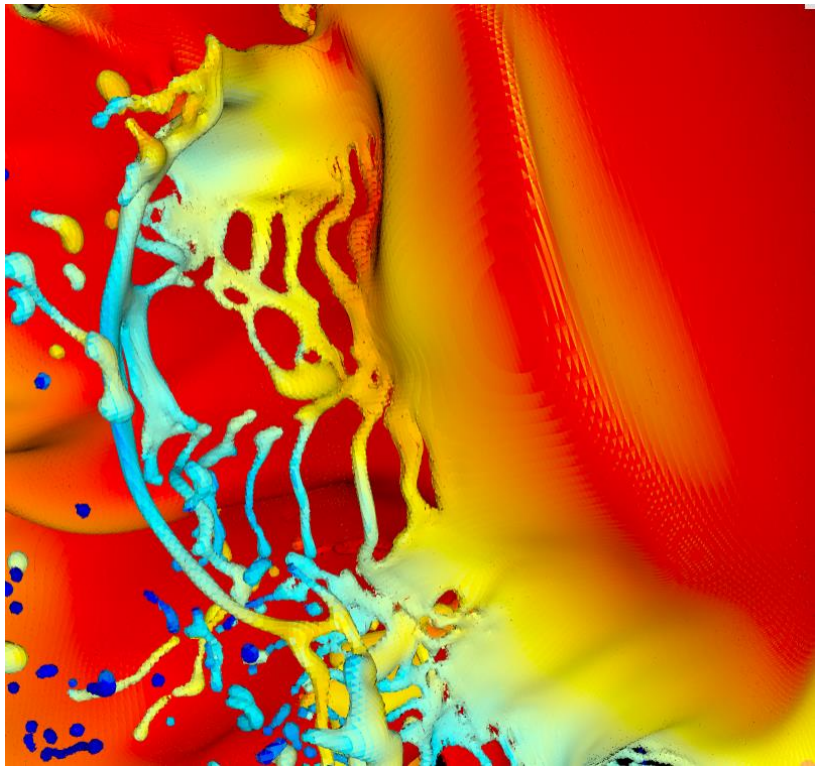

Figure 16. Structure of power-law jet tip and sheet thinning at t*=9. Jet surface is colored by axial velocity ($u$)

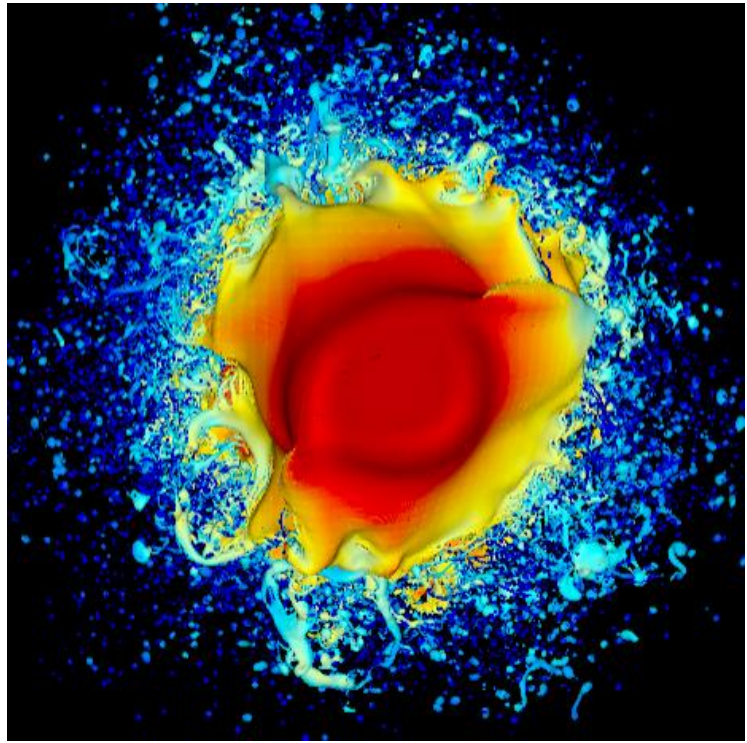

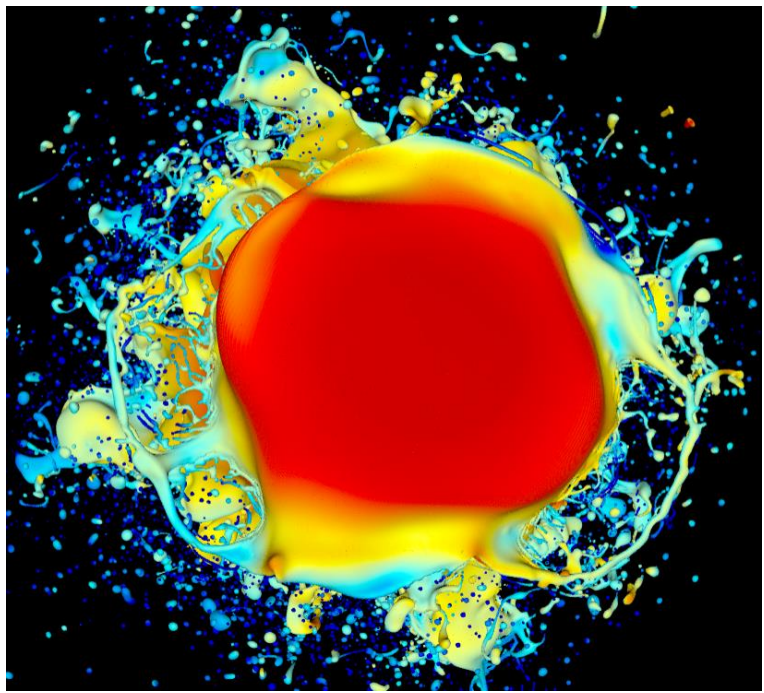

Figure 17. Comparison of jet tip for (a) parabolic and (b) Power law jets